\documentclass[%
 reprint,
 amsmath,amssymb,
 aps,
]{revtex4-2}

\usepackage{graphicx}% Include figure files
\usepackage{dcolumn}% Align table columns on decimal point
\usepackage{bm}% bold math
\begin{document}

\title{Equations of state and stability condition of mixed p-spin glass model}

\author{Ali Talebi}
 \email{ali.talebi.physics@gmail.com}
\affiliation{Department of Physics, Shahid Beheshti University, Evin, Tehran 1983969411, Iran}%
\date{\today}
%%%%%%%%%%%%%%%%%%%%%%%%%%%%%%%%%%%%%%%%%%%%%%%%%%%%%%%%%%%%%%%%%%%%%%%%
\begin{abstract}
The Sherrington-Kirkpatrick (SK) is a foundational model for understanding spin glass systems. It is based on the pairwise interaction between each two spins in a fully connected lattice with quenched disordered interactions. The nature of long-range interaction among spins in the (SK) model simplifies the study of this system by eliminating fluctuations. An advanced (SK) model, known as the p-spin model, introduces higher-order interactions that involve the interaction of P spins. This research focuses on the general Hamiltonian of the spin glass model with long-range interaction, referred to as the mixed p-spin glass model, which consists of adding all p-spin interaction terms. This research aims to derive the equation of states for this Hamiltonian, formulate the equation of state within the framework of the first replica symmetry breaking, and determine both the stability condition of the replica symmetric solution and the stability of the replicas belonging to the same group in the first step of replica symmetry breaking. 
\end{abstract}
%%%%%%%%%%%%%%%%%%%%%%%%%%%%%%%%%%%%%%%%%%%%%%%%%%%%%%%%%%%%%%%%%%%%%%%%%
\maketitle
\section{Introduction}
%paragraph 1 
The spin glass system is a magnetic system characterized by disordered interactions. In this system, the coexistence of positive and negative interactions leads to a phenomenon known as frustration. This frustration is the reason for many local minima of energy at low temperatures, making it challenging to explore the properties of this system. Studying this frustrated system has practical applications in different fields, including optimization\cite{kirkpatrick1984optimization}, quantum computing\cite{kadowaki1998quantum}, and the concept of learning\cite{amit1985storing, gardner1988optimal, gardner1988space}. The Edwards–Anderson model (EA)\cite{AndersonEdwards} was the first model introduced to examine the magnetic behavior of spin glass systems, where each spin interacts with its nearest neighbors. This short-range interaction, combined with frustration, makes the study of this model challenging. In contrast, the Sherrington-Kirkpatrick (SK) \cite{SK1} model featured a spin glass model with long-range interactions, where each spin has a pairwise interaction with other lattice spins. The exact solution of the SK model is equivalent to the mean field approximation, which makes it easier to study this system. 

%paragraph2
The SK model was initially solved under the replica symmetry (RS) assumption, which predicted the existence of spin glass, paramagnetic, and ferromagnetic phases. The RS solution accurately describes the system's transition from spin glass to paramagnetic phase and from ferromagnetic to paramagnetic phase. Nevertheless, this assumption fails to indicate the accurate value for the system's entropy at low temperatures and cannot determine the unstable region in the phase diagram. Almeida and Thouless \cite{Thouless_1978} found the stability condition for the RS solution, which in the phase diagram is represented as the AT line. Later, Parisi \cite{Parisi1, PARISI1979203, Parisi1980, Parisi3, Parisi4} introduced the replica symmetry-breaking (RSB) solution to find a new equation of state at low temperatures. This solution could solve the challenge of negative entropy at temperatures close to zero. By analyzing the replica symmetry-breaking solution of susceptibility, Parisi could find the exact boundary of the phase transition from the spin-glass to the ferromagnetic phase. The RSB solution is a mathematical framework for studying the correlation between different replicas. There are also various non-replica methods for investigating the spin glass system. Other non-replica methods, such as the TAP method \cite{TAP}, use the mean-field theory to study free energy variation, and the cavity method \cite{Spin_Glass_Theory_and_Beyond} is based on eliminating a spin from the system to explore the variation of the average local field.      

%paragraph3
The study of spin glass systems requires order parameters. Magnetization and overlap are order parameters that illustrate the system's phase and phase transition. In the spin glass system, the overlap determines whether the system is frozen or dynamic and characterizes the phase transition from the paramagnetic phase to the spin glass phase. The replica symmetric overlap is an autocorrelation measurement, while the replica symmetry-breaking overlap represents the cross-correlation between each pair of replicas \cite{mezard1984nature}. Order parameters play a role in the self-consistent equation of state, where the possibility of finding a solution by changing the temperature represents the phase transition. 

%paragraph4
The spin glass system with higher-order interaction, known as the p-spin model, which involves the interactions among P spins  \cite{gross1984simplest, GARDNER1985747}, has applications in image restoration and error-correcting codes \cite{nishimori1999statistical}. The special limit when $p \xrightarrow{} N$, and $N \xrightarrow{} \infty$ is also known as the simplest spin glass model \cite{PhysRevB.24.2613} without replica symmetry breaking at low temperature. In this article, I explore the general form of the spin glass Hamiltonian, known as the mixed p-spin glass model, where different p-spin interaction terms appear with distinct mean and variance. There is also some research to explore the Parisi formula of free energy for the mixed p-spin glass model \cite{talagrand2006parisi, panchenko2014parisi}. However, I aim to find the equations of state and the stability condition for the mixed p-spin glass Hamiltonian. There is research to explore the special form of the mixed p-spin model where the Sherrington-Kirkpatrick (SK) Hamiltonian is modified with the addition of a perturbative triadic interaction \cite{PhysRevE.109.064105}, which is limited to perturbative triadic interaction, and I expanded the findings from this research to a more general context. The analysis in this article is well known in the context of the spin glass system.
%%%%%%%%%%%%%%%%%%%%%%%%%%%%%%%%%%%%%%%%%%%%%%%%%%%%%%%%%%%%%%%%%%%%%%%%%%
\section{Model}\label{model}
This article investigates the general Hamiltonian for the spin glass model with long-range interaction, commonly known as the mixed p-spin model. The Hamiltonian 
\begin{equation}\label{EQ:1}
H=\sum_{p=1}^NH_p
\end{equation}
includes the addition of p-spin Hamiltonians as 
\begin{equation}\label{EQ:2}
 H_p = - \sum_{i_{1}<\ldots<i_{p}} J_{i_{1}\ldots i_{p}} \ S_{i_{1}} \ldots \ S_{i_{p}}
\end{equation}
In this model, \textbf{S} $ \in \{+ 1, -1\}$ represents lattice spin and the term $J_{i_{1}\ldots i_{p}}$ for $p\geq2$ denotes to quenched interaction among $p$ spins. Additionally, $J_i$ refers to the random external field, which is commonly known as $h_i$; however, it is changed to simplify the notation. All interaction terms follow a Gaussian Probability distribution by
\begin{equation}\label{EQ:3}
\begin{aligned}
&P\left(J_{i_1, \ldots, i_p}\right)=\\
&\sqrt{\frac{N^{p-1}}{\sigma_p^2 \pi p !}} \exp \left(\frac{-N^{p-1}}{\sigma_p^2 p !}\left(J_{i_1, \ldots, i_p}-\frac{\mu_p p !}{N^{p-1}}\right)^2\right)\\
\end{aligned}
\end{equation}
. The mean and variance of interactions need to be renormalized as
\begin{equation}\label{EQ:4}
\left[J_{i_1, \ldots, i_p}\right]=\frac{\mu_{p} p !}{N^{p-1}},\quad \left[\left(\Delta J_{i_1, \ldots, i_p}\right)^2\right]=\frac{\sigma_p^2 p !}{N^{p-1}}
\end{equation}
because the energy and specific heat are two extensive measurements that should be proportional to N. 
\section{Analysis}\label{Analysis}
To analyze the thermodynamic behavior of the system, it is essential to calculate the average of free energy over all possible realizations of interaction per degree of freedom. This procedure allows us to capture the effect of disordered in the system. This can be expressed by the following equation
\begin{equation}\label{EQ:5}
-\beta \mathbb{E} \left[f\right]=\frac{1}{N} \mathbb{E} \left[\ln{Z}\right]
\end{equation}
A well-known method for simplifying the calculation involves using the replica trick \cite{SK1}. The idea is to first calculate $\mathbb{E}\left[Z^{n}\right]$ for integer n, eventually taking the limit $n \rightarrow 0$, instead of calculating $ \mathbb{E} \left[ \ln{Z} \right] $ directly. This approach is formulated as
\begin{equation}\label{EQ:6}
\mathbb{E}\left[\ln{Z}\right]=\lim_{n \rightarrow 0} \frac{\mathbb{E}\left[Z^n\right]-1}{n}
\end{equation}
, the rest of the calculation is based on the formal analysis of the p-spin model developed by Gardner \cite{GARDNER1985747}. First, it needs to take the average over the interaction parameter based on the Gaussian integral of the probability distribution function to calculate the average partition function. This average is taken by \\
\begin{widetext}
\begin{equation}\label{EQ:7}
\begin{aligned}
& {\mathbb{E} \left[Z^n\right]= \int\left(\prod_{ p=1 }^{N}\prod_{i_1 <\ldots<i_p } dJ_{i_1 \ldots i_p} p\left(J_{i_1 \ldots i_p}\right)\right) \cdot \operatorname{Tr} \exp \left(\beta \sum_{p=1}^N \sum_{i_1 <\ldots<i_p} J_{i_1 \ldots i_p} \sum_{\alpha=1}^n S_{i_1}^\alpha \ldots S_{i_p}^\alpha\right)} \\
\end{aligned}
\end{equation}
Applying the transformation $J_{i_1 \ldots i_p}-\frac{\mu_{p} p !}{N^{p-1}} \rightarrow J_{i_1 \ldots i_p}$ in evaluation of gaussian integral change the expression $\mathbb{E}\left[Z^n\right]$ as
\begin{equation}\label{EQ:8}
{\mathbb{E}\left[Z^n\right]} =\operatorname{Tr} \exp \left\{\sum_{p=1}^{N}\frac{\beta^2 \sigma_p^2 N}{2} \sum_{\alpha<\beta}\left(\frac{1}{N} \sum_i S_i^\alpha S_i^\beta\right)^p+\sum_{p=1}^N\frac{\beta^2 \sigma_p^2}{4} N n+\sum_{p=1}^{N}\mu_{p} \beta N \sum_\alpha\left(\frac{1}{N} \sum_i S_i^\alpha\right)^p\right\}
\end{equation}
The next step in calculation needs to define overlap ($q_{\alpha\beta}=\frac{1}{N}\sum_i\sigma_i^\alpha\sigma_i^\beta$), and magnetization ($m_\alpha=\frac{1}{N}\sum_i\sigma_i^\alpha$) as two order parameters. By employing the Fourier transform of the delta function, both order parameters and their conjugate variables appear in the expression $\mathbb{E}\left[Z^n\right]$ as follow 
\begin{equation}\label{EQ:9}
\begin{aligned}
{\mathbb{E}\left[Z^n\right]=} & \operatorname{Tr} \int \prod_{\alpha<\beta} \mathrm{d} q_{\alpha \beta} \mathrm{d} \hat{q}_{\alpha \beta} \int \prod_\alpha \mathrm{d} m_\alpha \mathrm{d} \hat{m}_\alpha \exp \left\{\sum_{p=1}^{N} \frac{\beta^2 \sigma_p^2 N}{2} \sum_{\alpha<\beta}\left(q_{\alpha \beta}\right)^p\right.  -N \sum_{\alpha<\beta} q_{\alpha \beta} \hat{q}_{\alpha \beta}+N \sum_{\alpha<\beta} \hat{q}_{\alpha \beta}\left(\frac{1}{N} \sum_i S_i^\alpha S_i^\beta\right)\\
&+\sum_{p=1}^{N} \mu_{p} \beta N \sum_\alpha\left(m_\alpha\right)^p 
 \left.-N \sum_\alpha m_\alpha \hat{m}_\alpha+N \sum_\alpha \hat{m}_\alpha\left(\frac{1}{N} \sum_i S_i^\alpha\right)+\sum_{p=1}^N \frac{1}{4} \beta^2 \sigma_p^2 N n\right\}
\end{aligned}
\end{equation}
After applying the $\operatorname{Tr}$ operation on the expression, it changes to 
\begin{equation}\label{EQ:10}
\begin{aligned}
& {\mathbb{E}\left[Z^n\right]=\int \prod_{\alpha<\beta} \mathrm{d} q_{\alpha \beta} \mathrm{d} \hat{q}_{\alpha \beta} \int \prod_\alpha \mathrm{d} m_\alpha \mathrm{d} \hat{m}_\alpha \exp \left\{\sum_{p=1}^{N}\frac{\beta^2 \sigma_p^2 N}{2} \sum_{\alpha<\beta}\left(q_{\alpha \beta}\right)^p\right.}-N \sum_{\alpha<\beta} q_{\alpha \beta} \hat{q}_{\alpha \beta}+\sum_{p=1}^N\frac{1}{4} \beta^2 \sigma_p^2 N n+\sum_{p=1}^{N} \mu_{p} \beta N \sum_\alpha\left(m_\alpha\right)^p \\
& \left.-N \sum_\alpha m_\alpha \hat{m}_\alpha+N \log \operatorname{Tr} \exp \left(\sum_{\alpha<\beta} \hat{q}_{\alpha \beta} S^\alpha S^\beta+\sum_\alpha \hat{m}_\alpha S^\alpha\right)\right\}
\end{aligned}
\end{equation}
The replica-symmetric (RS) solution is under the assumption that different replicas are not distinguishable, which makes it necessary to replace $q_{\alpha \beta}$ with $q$. In the thermodynamic limit ($N \rightarrow \infty$), calculating the integral needs to use the steepest descent method. The result is
\begin{equation}\label{EQ:11}
\begin{aligned}
& {\mathbb{E}\left[Z^n\right] \approx \exp \left[N \left\{\sum_{p=1}^{N}\beta^2 \sigma_p^2 \frac{n(n-1)}{4} q^p-\frac{n(n-1)}{2} q \hat{q}+\sum_{p=1}^{N} \mu_p \beta n m^p-n m \hat{m}+\sum_{p=1}^N\frac{1}{4} n \beta^2 \sigma_p^2\right.\right.} \\
& \left.\left.\quad+\log \operatorname{Tr} \int \mathrm{D} u \exp \left(\sqrt{\hat{q}} u \sum_\alpha \sigma^\alpha+\hat{m} \sum_\alpha \sigma^\alpha-\frac{n}{2} \hat{q}\right)\right\}\right]
\end{aligned}
\end{equation}
and by replacing the expression $\mathbb{E}\left[Z^n\right]$ in the relation $\mathbb{E}\left[\ln{Z}\right]=\lim_{n \rightarrow 0} \frac{\mathbb{E}\left[Z^n\right]-1}{n}$, the free energy is
\begin{equation}\label{EQ:12}
-\beta \mathbb{E} \left[f\right]= -\sum_{p=1}^N\frac{1}{4} \beta^2 \sigma_p^2 q^p+\frac{1}{2} q \hat{q}+\sum_{p=1}^N\beta \mu_p m^p-m \hat{m} +\sum_{p=1}^N\frac{1}{4} \beta^2 \sigma_p^2-\frac{1}{2} \hat{q}+\int \mathrm{D} u \ln 2 \cosh (\sqrt{\hat{q}} u+\hat{m})
\end{equation}
The equations of state appear by variation of free energy with respect to the order parameters as
\begin{equation}\label{EQ:13}
\hat{q}=\sum_{p=1}^N\frac{1}{2} p \beta^2 \sigma_p^2 q^{p-1}, \quad \hat{m}=\sum_{p=1}^N\beta \mu_p p m^{p-1}
\end{equation}

\begin{equation}\label{EQ:14}
q=\int \mathrm{D} u \tanh ^2(\sqrt{\hat{q}} u+\hat{m}), \quad m=\int \mathrm{D} u \tanh (\sqrt{\hat{q}} u+\hat{m})
\end{equation}
 which needs to substitute $\hat{q}$ and $\hat{m}$ from Eq.(\ref{EQ:13}) in Eq.(\ref{EQ:14}) to eliminate these variables from equations of state. The equations for $q$ and $m$ with real variables can be written as
\begin{equation}\label{EQ:15}
q=\int \mathrm{D} u \tanh ^2 \beta G, \quad m=\int \mathrm{D} u \tanh \beta G
\end{equation}
where $G$ is determined by
\begin{equation}\label{EQ:16}
G= \sqrt{\frac{\sum_{p=1}^N\sigma_p^2p q^{p-1}}{2}} u+\sum_{p=1}^N\mu_p p m^{p-1}
\end{equation}
that is the equations of state concerning the replica symmetric assumption. The first step of replica symmetry-breaking modifies the free energy of Eq.(\ref{EQ:12}) as
\begin{equation}\label{EQ:17}
\begin{aligned}
& -\beta \mathbb{E}\left[f\right]=-\hat{m} m+\frac{1}{2} x \hat{q}_0 q_0+\frac{1}{2}(1-x) \hat{q}_1 q_1+\sum_{p=1}^N\beta \mu_p m^p  -\sum_{p=1}^N\frac{1}{4} x \beta^2 \sigma_p^2 q_0^p-\sum_{p=1}^N\frac{1}{4}(1-x) \beta^2 \sigma_p^2 q_1^p+\sum_{p=1}^N\frac{1}{4} \beta^2 \sigma_p^2-\frac{1}{2} \hat{q}_1 \\
& +\frac{1}{x} \int \mathrm{D} u \ln \int \mathrm{D} v \cosh ^x\left(\hat{m}+\sqrt{\hat{q}_0} u+\sqrt{\hat{q}_1-\hat{q}_0} v\right)+\ln 2
\end{aligned}
\end{equation}
where two additional order parameters, $q_1$ and $\hat{q_1}$, are added to the free energy, leading to new equations of state as follows
\begin{align}
m & =\int \mathrm{D} u \frac{\int \mathrm{D} v \cosh ^x \beta G_1 \tanh \beta G_1}{\int \mathrm{D} v \cosh ^x \beta G_1} \label{EQ:18} \\
q_0 &\int \mathrm{D} u\left(\frac{\int \mathrm{D} v \cosh ^x \beta G_1 \tanh \beta G_1}{\int \mathrm{D} v \cosh ^x \beta G_1}\right)^2 \label{EQ:19} \\
q_1 & =\int \mathrm{D} u \frac{\int \mathrm{D} v \cosh ^x \beta G_1 \tanh ^2 \beta G_1}{\int \mathrm{D} v \cosh ^x \beta G_1} \label{EQ:20} \\
G_1 & = \sqrt{\frac{\sum_{p=1}^N\sigma_p^2 p q_0^{p-1}}{2}} u+ \sqrt{\sum_{p=1}^N \sigma_p^2\frac{p}{2}\left(q_1^{p-1}-q_0^{p-1}\right)} v+\sum_{p=1}^N \mu_p p m^{p-1} \label{EQ:21}
\end{align}
\end{widetext}
The stability condition of the replica symmetric solution is written as the Almeida and Thouless method \cite{Thouless_1978}. To find the replica symmetry stability condition, the expansion of free energy needs to be investigated to second order, where the positive eigenvectors of the Hessian matrix represent the stability of the replica symmetric solution. The replica symmetric stability condition is as follows
\begin{equation}\label{EQ:22}
\frac{2 T^2 \sum_{p=1}^N \sigma_p^2 p(p-1)q^{p-2}}{(\sum_{p=1}^N \sigma_p^2p(p-1)q^{p-2})^2}> \int \mathrm{D} u \operatorname{sech}^4 \beta G
\end{equation}
The stability condition for first replica symmetry breaking, when the indices $\alpha$ and $\beta$ of $q_{\alpha\beta}$ belong to the same block, is 
\begin{equation}\label{EQ:23}
\frac{2 T^2 \sum_{p=1}^N \sigma_p^2 p(p-1)q_1^{p-2}}{(\sum_{p=1}^N \sigma_p^2p(p-1)q_1^{p-2})^2} > \int \mathrm{D} u \frac{\int \mathrm{D} v \cosh ^{x-4} \beta G_1 }{\int \mathrm{D} v \cosh ^x \beta G_1}
\end{equation}
that represents the stability of the replica symmetric solution at the first replica symmetry breaking of the mixed p-spin Hamiltonian.
%%%%%%%%%%%%%%%%%%%%%%%%%%%%%%%%%%%%%%%%%%%%%%%%%%%%%%%%%%%%%%%%%%%%%%
\section{Conclusion}
In this work, I have explored the free energy of the mixed p-spin glass Hamiltonian, which enabled me to find the replica symmetric equations of state and the first replica symmetry-breaking equations of state referred to this Hamiltonian. Moreover, the analysis of the stability condition for the replica symmetric solution, represented in Eq.(\ref{EQ:22}), reveals that the external field term ($p=1$) does not influence the stability of the replica symmetric solution in the mixed p-spin glass Hamiltonian. The same conclusion holds for the stability of the first step replica symmetry-breaking solution Eq.(\ref{EQ:23}) when the replicas belong to the same group.
This article focused on the mathematical foundation of the mixed p-spin glass model. Despite significant theoretical progress, several open problems remain concerning the application of this model with higher-order interactions in fields such as neuroscience and social science.   

\bibliographystyle{unsrt}
\bibliography{apssamp}

\end{document}